\documentclass[twocolumn]{optica-article}

\journal{opticajournal} 

\articletype{Research Article}
\usepackage{float}
\usepackage{lineno}
\usepackage{soul}
\usepackage{acronym}
\usepackage{tabularx}
\usepackage{multicol}

\begin{document}

\newacro{SOI}{Silicon-on-Insulator}
\newacro{WGM}{Whispering Gallery Mode}
\newacro{EBPG}{Electron-Beam Pattern Generator}
\newacro{EBL}{Electron-Beam Lithography}
\newacro{RIE}{Reactive-Ion Etching}
\newacro{IPA}{isopropyl alcohol}
\newacro{HF}{hydroflouric acid}
\newacro{VPE}{Vapor Phase Etching}
\newacro{FSR}{Free Spectral Range}
\newacro{SNR}{Signal-to-Noise Ratio}
\newacro{PEC}{Proximity Effect Correction}
\newacro{SEM}{Scanning Electron Microscope}
\newacro{FEM}{Finite Element Method}
\newacro{Si}{silicon}
\newacro{SiO$_{2}$}{silica}
\newacro{PD}{photodector}
\newacro{PSD}{Power Spectrum Density}
\newacro{FWHM}{Full Width at Half Maximum}
\newacro{DRIE}{Deep Reactive Ion Etching}
\newacro{BOX}{buried oxide}
\newacro{PC}{Polarization Controller}
\newacro{HWHM}{Half Width at Half Maximum}

\title{Silicon Double-Disk Optomechanical Resonators from Wafer-Scale Double-Layered Silicon-on-Insulator}

\author{Amy Navarathna,\authormark{1,2} Benjamin J. Carey,\authormark{1,2} James S. Bennett,\authormark{3} Soroush Khademi,\authormark{1,2} and Warwick P. Bowen\authormark{1,2,4*}}

\address{\authormark{1}School of Mathematics and Physics, The University of Queensland, St Lucia, Queensland 4072, Australia.\\
\authormark{2}ARC Centre of Excellence for Engineered Quantum Systems, St Lucia, Queensland 4072, Australia.\\
\authormark{3}Centre for Quantum Dynamics, Griffith University, Nathan, Queensland 4222, Australia.\\
\authormark{4}ARC Centre of Excellence in Quantum Biotechnology, St. Lucia, Queensland 4072, Australia\\}
\email{\authormark{*}w.bowen@uq.edu.au} 


\begin{abstract*} 
\ac{WGM} optomechanical resonators are a promising technology for the simultaneous control and measurement of optical and mechanical degrees of freedom at the nanoscale. They offer potential for use across a wide range of applications such as sensors and quantum transducers. Double-disk \ac{WGM} resonators, which host strongly interacting mechanical and optical modes co-localized around their circumference, are particularly attractive due to their high optomechanical coupling. Large-scale integrated fabrication of silicon double-disk \ac{WGM} resonators has not previously been demonstrated. In this work we present a process for the fabrication of double-layer silicon-on-insulator wafers, which we then use to fabricate functional optomechanical double silicon disk resonators with on-chip optical coupling. The integrated devices present an experimentally observed optical quality factors of the order of $10^5$ and a single-photon optomechanical coupling of approximately 15~kHz.
\end{abstract*}

\section{Introduction}
\label{Sec:DDintroduction}	
Optomechanical \ac{WGM} resonators have applications in many fields including precision sensors \cite{Frustaci_WGM_2019, Liu_sensors_2021, Gotardo_23, basiri-esfahani_precision_2019}, biosensors \cite{Frustaci_WGM_2019}, memory \cite{kristensen2023longlived}, communications \cite{lei_fully_2022} and for quantum information devices \cite{ chan_optical_2009, bowen_quantum_2015, Tang_entanglment_2022}. Such resonators support optical modes confined to their periphery by the refractive index contrast at their outer boundary. They also support mechanical degrees of freedom that strongly overlap spatially with the confined optical field \cite{mancini_optomechanical_1998, aspelmeyer_cavity_2014}, leading to significant optomechanical coupling. \ac{WGM} optomechanical resonators have been experimentally demonstrated in structures such as microspheres \cite{ma_radiation-pressure-driven_2007,shen_reconfigurable_2018}, microtoroids \cite{lee_cooling_2010,verhagen_quantum-coherent_2012,ruesink_optical_2018,bekker_free_2018, bekker2017injection}, (single) microdisks \cite{lu_high-frequency_2015,jiang_chip-scale_2016}, and double-disks \cite{jiang_high-q_2009,meng_measurement-based_2022,lei_fully_2022}.

A double-disk \ac{WGM} resonator is made by fabricating two coaxial disks separated by a narrow air gap. Evanescent coupling between the two disks results in hybridization of the \ac{WGM}s, forming modes that have a high field intensity in the gap\cite{jiang_high-q_2009}. This is illustrated in Fig.~\ref{fig:modeling}(a). Modulation of the air gap by out-of-plane mechanical motion results in large optomechanical coupling; particularly for the symmetrical out-of-plane oscillation (as shown in Fig.~\ref{fig:modeling}(b)). This is compounded by the tendency for out-of-plane deformations to be much larger than in-plane deformations (of equal energy), due to bending typically requiring less energy than stretching or compressing a material \cite{ward2011wgm}. As such, double-disk systems achieve some of the largest optomechanical coupling rates in \ac{WGM} optomechanical devices demonstrated to date \cite{meng_measurement-based_2022}.

The optomechanical coupling can also be increased by improving the confinement of the optical mode and ensuring strong co-localization of the optical and mechanical modes\cite{chan_optical_2009,chan_laser_2011,gavartin_optomechanical_2011,leijssen_nonlinear_2017,ren_two-dimensional_2020,lu_silicon_2020}.
Both objectives can be achieved by moving to smaller resonators fabricated from materials with higher refractive indices. Double-disk optomechanical resonators have previously been fabricated from materials such as silica (SiO$_{2}$, $n\approx 1.44$) \cite{meng_measurement-based_2022, jiang_high-q_2009, bekker_free_2018, lei_fully_2022, lin2009mechanical},  silicon nitride (Si$_3$N$_4$, $n\approx 1.98$) \cite{lee2010silicon, wiederhecker2011broadband}, and lithium niobate (LiNO$_\text{3}$, $n\approx 2.21$)    \cite{zheng2019high}. There is much interest in developing \ac{Si} optomechanical devices for many applications \cite{moradinejad_double-layer_2014,espinel_brillouin_2017,dehghannasiri_integrated_2018}. Silicon's larger refractive index ($n=3.48$) permits tighter confinement of optical modes and hence higher coupling rates and further miniaturization of devices. This potential is further bolstered by silicon's compatibility with electronics fabrication processes and operation with light in the telecommunications C-band, leading to opportunities for integration with a range of technologies and systems. In particular, \ac{Si} optomechanical resonators present ready compatibility with silicon photonic, semiconducting, and superconducting circuits. Further, recent studies have demonstrated control of defect-induced single photon emissions in silicon substrates \cite{Hollenbach:2020-defect, Durand-2021_singe-photon-defect} offering the potential for integration with quantum information systems utilizing such photon sources. 

Previous reports have demonstrated chip-scale bonding to produce high-purity Si double-layers with the necessary oxide spacing layer ($\sim 100$~nm) for production of free-standing Si double-disks \cite{moradinejad_double-layer_2014} or oxide-clad ring resonators \cite{briggs_wafer-bonded_2009}. However, due to the complicated nature of their production, demonstration of large-scale Si double-disk WGM resonator fabrication and experimental investigations of their optomechanical performance remained elusive.

In this work we present a truly wafer-scale approach for double layer \ac{SOI} substrate fabrication, with subsequent chip-scale fabrication and characterization of free-standing double-silicon-disk WGM optomechanical resonators. We achieve this by bonding two entire 4-inch SOI wafers prior to device fabrication. This allows for both chip- and wafer-scale double-layer~SOI photonics, which are compatible with the aforementioned technologies. Further, we present \ac{FEM} modeling and comparison to experimental results of the fabricated devices. We observe optical and mechanical resonances of the disks under vacuum, and characterize the single-photon optomechanical coupling rate ($g_0$) through measurements of the optomechanical spring effect.

\section{Device Design and Fabrication}
\label{results:FEM}
\begin{figure}[bht]
    \centering
    \includegraphics[width=0.95\textwidth]{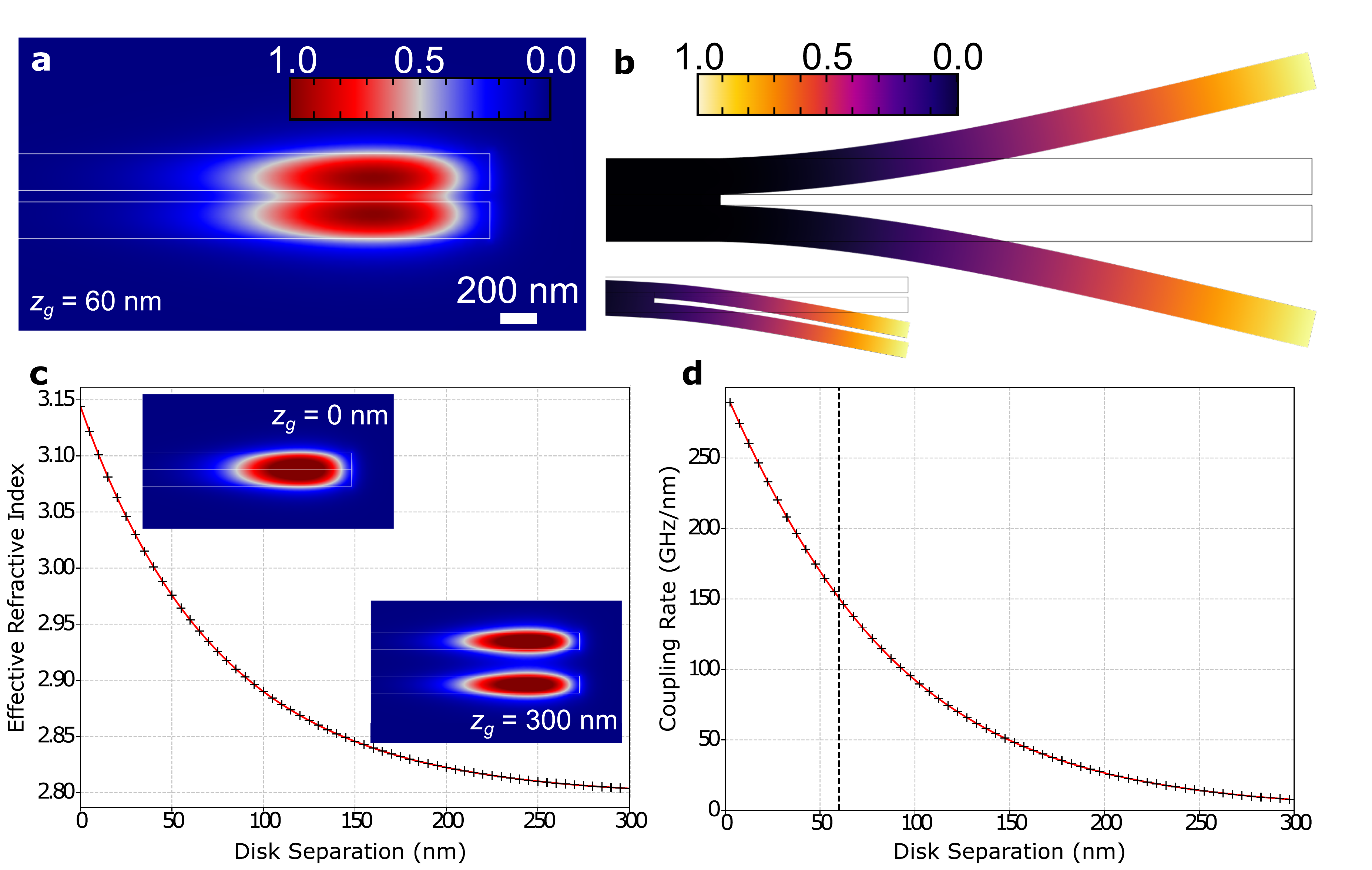}
    \caption[Finite-element-method simulation results for silicon double-disks.]{Axisymmetric eigenmode \ac{FEM} simulation results of the double-disk resonators.\\
    (a) The optical field distribution of a hybridized (transverse electric) whispering gallery mode.\\
    (b) Mechanical displacement profile of the symmetric mechanical resonance (inset: asymmetric mode-profile).\\
    (c) Effective refractive index of the optical mode (\textit{cf}. Eq.~\ref{eq.neff}) with differing disk separation (insets: mode profiles with 0~nm \& 300~nm spacings).\\
    (d) Optomechanical coupling rate of the disks for varying separations (the vertical line corresponds to the 60~nm used in our fabricated devices).}
    \label{fig:WGM}
    \label{fig:modeling}
\end{figure}

As described above, double-disk resonators consist of two coaxial \ac{Si} microdisks separated vertically by a \ac{SiO$_{2}$} sandwich layer of thickness $z_g$, acting as a spacer. This \ac{SiO$_{2}$} layer is selectively etched around the periphery, resulting in an air gap between the disks. When the gap is sufficiently small, evanescent coupling results in the emergence of a hybridized optical mode that exists in both disks and the gap simultaneously; this is depicted in Fig.~\ref{fig:modeling}(a). As expected, our optical FEM simulation (COMSOL Multiphysics) shows that the fundamental (transverse) \ac{WGM}s are confined to the outer $\sim 2~\upmu\mathrm{m}$ of the disks, which is commensurate with the region of greatest deformation caused by mechanical modes. The fundamental symmetric flapping mode is shown in Fig.~\ref{fig:modeling}(b).

To characterize the impact of the motion on the optical \ac{WGM}, we first consider the effective refractive index ($n_\text{eff}$), which can be determined from the weighted integral method\cite{ohke1995new}:
\begin{equation}
    	\label{eq.neff}
     n_\text{eff} = \left(\frac{\int_V E(\vec{r}) n^2(\vec{r}) \;\mathrm{d}V}{\int_V E(\vec{r}) \;\mathrm{d}V}\right)^{\frac{1}{2}}.
\end{equation}
Here, $E$ and $n$ are the electric field amplitude and material refractive index, respectively, at position $\vec{r}$ within the simulation volume $V$. The effective index of the \ac{WGM}s with varying disk separation is presented in Fig.~\ref{fig:modeling}(c). When there is no air-gap (\textit{i.e.} $z_g = 0$) the WGM behaves as if contained within a single 400-nm disk, with the field well-confined inside the Si; thus, $n_\text{eff}$ approaches $n_\text{Si}$. Conversely, when the gap is very large (\textit{e.g.} $z_g \geq 300$~nm) the mode resembles two independent WGMs with negligible shared field (see Fig.~\ref{fig:modeling}(c) insets). The effective refractive index is lower than that of the 0-nm gap because a larger fraction of the optical field extends evanescently beyond the Si.

The angular resonance frequencies of an optical \ac{WGM} ($\omega_m$) with a radius of $r$ are approximately given by $\omega_{m} = m{c}/{r n_{\text{eff}}}$, where $m$ is the mode number associated with the number of wavelengths within the cavity and $c$ is the speed of light in vacuum. The dependence of $\omega_m$ on $n_{\text{eff}}$, which is modulated by mechanical motion, ensures that changes in the disk separation will result in a shift in optical resonance frequency. For small deformations, the dependence can be summarised by the optomechanical coupling  (we suppress the subscript for the $m$th mode),
\begin{equation}
    G_\text{OM} = \frac{\delta \omega_m}{\delta z_g}.
\end{equation}
A high $G_\text{OM}$ value is advantageous for applications such as sensing and transduction \cite{Xia_Optomech_review_2020, lauk_perspectives_2020}. Our FEM model is presented in Fig.~\ref{fig:modeling}(d). Here, it can be seen that the optomechanical coupling diminishes with increasing separation. The maximum coupling rate occurs as $z_g $ approaches zero; however, this regime is experimentally impracticable due to large van der Waal's and Casimir forces that can promote collapse of the disks, effectively destroying the mechanical modes of interest \cite{rodriguez2011designing, rodriguez2011bonding}. For these reasons, and to ensure fabrication consistency, a separation of 60~nm was chosen. With a disk separation of 60~nm we predict an optomechanical coupling of $G_\text{OM}/2\pi\approx$150~GHz/nm, which is among the highest reported in optomechanical cavities \cite{Winger_2011, Xia_Optomech_review_2020, woolf2013optomechanical}.

A second key figure of merit is the single-photon optomechanical coupling rate, which 
is given by:
\begin{equation}
    g_0 = x_{\text{zpf}}G_{\text{OM}},
\end{equation}
where $x_{\text{zpf}} = \sqrt{{\hbar}/{2m_{\text{eff}}\Omega}}$ is the zero point motion of the oscillation with resonance frequency of $\Omega/2\pi$, $\hbar$ is the reduced Planck's constant, and
\[
    m_{\text{eff}} =\int_{V} \rho(\vec{r}) \frac{\left|\vec{X}(\vec{r})\right|^2}{\left| \vec{X}_{\text{max}}\right|^2 } \,\mathrm{d}V
\] 
is the effective mass of the mechanical oscillation. Here $\rho$ is the material's density, $\vec{X}$ is the displacement, and $\vec{X}_{\text{max}}$ is the maximum displacement amplitude of the excitation. The single-photon coupling rate is equal to the frequency shift per zero-point displacement of the oscillator, and also determines the optical force per photon \cite{aspelmeyer_cavity_2014}.

Using the data extracted from both optical and mechanical FEM simulations (see S.1 for the mechanical FEM), a 120~$\upmu$m radius Si double--disk resonator with an air gap of 60~nm and undercut of 3.35~$\upmu$m yields a single photon coupling rate of $g_0/2\pi = 116~\text{kHz}$. 

\subsection{Further design considerations}

The design of our double-disk devices broadly follows the concepts of those presented in \cite{moradinejad_double-layer_2014}. However, unlike previous reports---which demonstrated only chip-scale double-layer~SOI---we present a wafer-scale fabrication process utilizing two whole wafers of commercially available \ac{SOI} (see Sec.~\ref{Sec:DDfab} \& S.2.1). Furthermore, unlike many previous works that have coupled light directly from tapered optical fibers into the \ac{WGM} cavity \cite{jiang_high-q_2009, meng_measurement-based_2022, bekker_free_2018}, our work utilizes coupling via integrated on-chip waveguides, enabling superior flexibility and scalability.

Our wafer-scale double-layer~SOI approach allows for rapid production of many devices, allowing for a range of devices of varying radii and waveguide-to-disk coupling distance to be fabricated and explored on the same chip. The radii of our devices range from 15~$\upmu$m to 120~$\upmu$m. A minimum radius of 15~$\upmu\mathrm{m}$ is used to ensure radiative optical losses in the devices are negligible \cite{buck_joseph_robert_cavity_2003}. This range of radii presents free spectral ranges (FSRs) in the range of 1--7~nm, allowing for expedition of experimental probing.

The coupling distance ranges from 100--400~nm. The on-chip waveguides are 500~nm wide, which ensures single mode operation around 1550~nm wavelength. The waveguides' two right-angled bends form an S-shaped configuration, minimizing the potential for stray light (\textit{e.g.}, light reflected off the substrate) to couple into the output. The waveguides are tapered down to 100~nm over the course of 50~$\upmu$m, allowing for mode-matched coupling to and from tapered optical fiber tips in a similar manner to that presented in \cite{Tiecke2015}. Thin tethers along the length of the waveguides minimize buckling due to stress in the Si films. These tethers are narrower than the waveguide to suppress their scattering cross-section.


\section{Fabrication}
\label{Sec:DDfab}

\begin{figure}[!ht]
    \centering
    \includegraphics[width=1\textwidth]{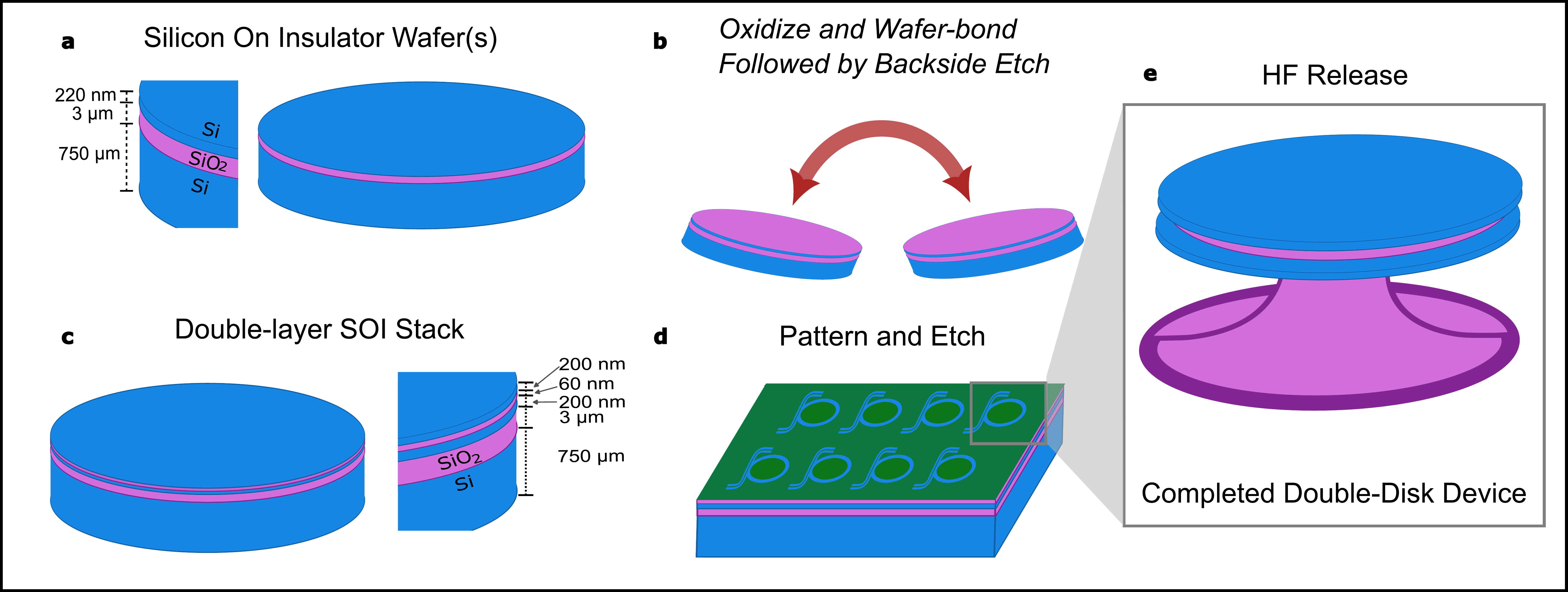}
     \caption{Wafer scale production of the Si double-disks from two standard silicon-on-insulator (SOI) wafers (not to scale).\\
     (a) Initial SOI wafer stacks.\\
     (b) Oxidation of the top layers prior to wafer bonding, followed by backside etching from one side to remove one carrier and buried oxide layer (detail in Fig.~S2).\\
     (c) The resultant double-layer SOI wafer.\\
     (d) The wafers are then diced into 15~mm~$\times$~15~mm chips and patterned with electron beam lithography, etched and undercut to create the devices (detail in Fig.~S4).\\
     (e) A released double-disk resonator.\\
     Blue, Si; magenta, silica; green, electron-beam resist.
    }
    \label{fig:Fig_1}
\end{figure}

\subsection{Double-layer silicon-on-insulator wafer fabrication}

Wafer-scale fabrication of double-disk devices requires double-layer~SOI wafers. However, such wafers are not commercially available and so were produced in-house in a similar---though scaled-up---manner to that which is presented in~\cite{moradinejad_hybrid_2017}. Here \ac{SOI} wafers (4 inch) consisting of a 215-nm~upper layer of crystalline Si, a 3-$\upmu$m buried oxide (SiO$_{2}$) layer, and a 750-$\upmu$m~Si carrier layer were purchased from WaferPro. The wafers were thermally oxidized in a dry oxidation furnace (HiTech benchtop oxidation furnace) at 900$~^\circ\mathrm{C}$ for two hours, forming an SiO$_{2}$ layer at the surface of the wafers. This oxide layer forms one half of the spacer layer in the final double-disk devices, as depicted in Fig. \ref{fig:Fig_1}(b \& c).  Thin-film reflectometry thickness analysis (SCI FilmTek 2000m) revealed an oxide thickness of 30~nm and underlying silicon layer thickness reduced to 200~nm. This increase in overall thickness is consistent with prior work \cite{zhang_fundamentals_2019}, and occurs because of silica's lower density than Si. The oxidized SOI wafers were then inspected for cleanliness with darkfield microscopy; pristine surfaces are a requirement for successful wafer bonding. After determining the surfaces were free of contaminants, the wafers were surface activated by exposure to oxygen plasma (200~sccm, 50~W, Oxford Instruments PlasmaPro80). The bonding process was then performed using an EVG520~IS Hot Embosser. The process consists of two pistons being pushed together. Mounted onto the top piston is a piece of quartz and the bottom piston is flat steel. When the process starts the temperature is slowly ramped up to 600$^\circ~$C and a force of 3.1~kN is applied. The high temperature reduces the buildup of stress between the wafers, and increases the bonding strength \cite{moriceau_overview_2011}. Bond strength was then assessed by a razor blade test, as described by \cite{pantzas_measuring_2020}.

Following the bonding, one side of the stack is exposed to \ac{HF} \ac{VPE} in order to remove any $\mathrm{SiO}_2$ grown on the back of the wafer during the thermal oxidation process (as the underside of the wafers were also exposed in the oxidation furnace).
Afterwards, \ac{DRIE} (Plasma-Therm Versaline \ac{DRIE}) is used to remove the 750-$\upmu \mathrm{m}$ silicon handle layer on one side of the bonded stack. Here a two-step Bosch process is used, consisting of $\mathrm{C}_4\mathrm{F}_8$ gas, which forms a passivating layer, followed by an isotropic Si etch with $\mathrm{SF}_6$. This process was repeated and the progress of the etch was monitored periodically by removing the wafer stack and visually inspecting. Photographs of the etch progression throughout the DRIE process can be found in the Supplementary Material (Sec.~S.2, Fig.~S3), showing non-homogeneous removal of material. As a result, a final liquid \ac{HF} etch, used to remove the 3-$\upmu \mathrm{m}$ \ac{BOX} layer that is now partially exposed, leaves approximately 60\% of the wafer area usable for fabrication of devices. Finally, the double-layer~SOI substrate was diced into chips of 20~mm$~\times$~20~mm for the subsequent fabrication of optomechanical silicon double-disks.

\begin{figure}
    \centering
    \includegraphics[width=0.95\linewidth]{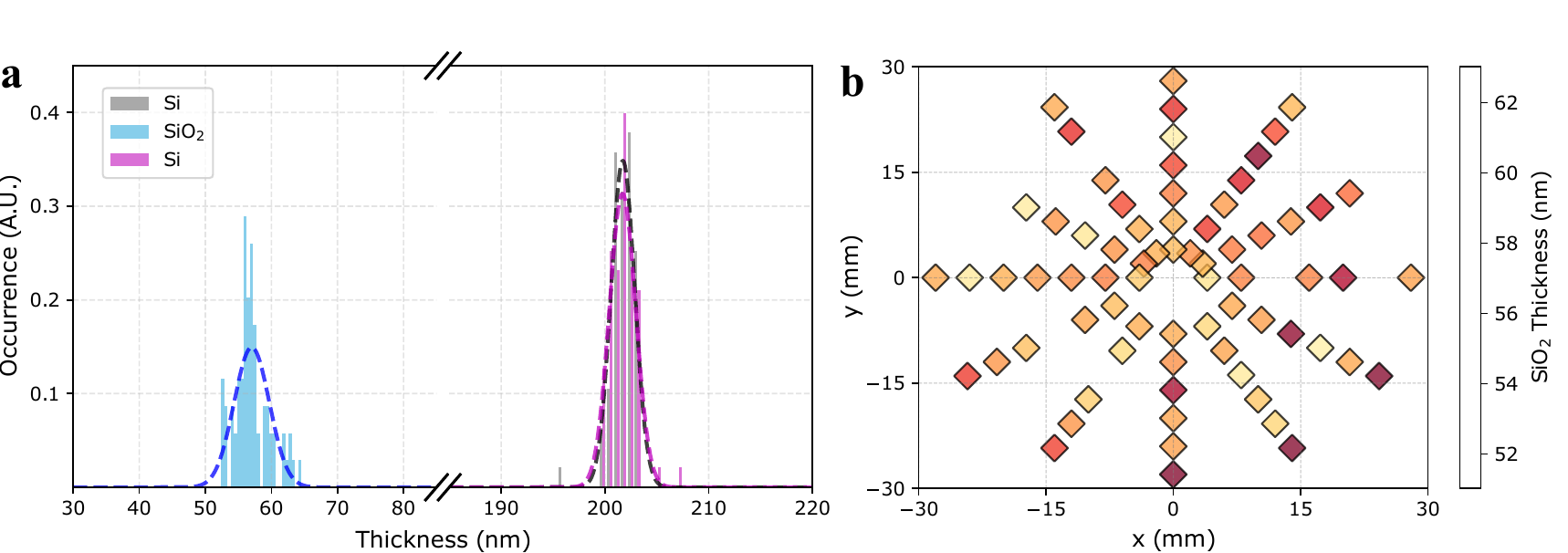}
    \caption{Assessment of the thickness uniformity of the double-layered SOI wafers determined with optical reflectometry. (a) Histograms depicting of the measured thickness distribution in the Si and SiO$_2$ layers with normal probability density functions fitted to the data. (b) Thickness map of the SiO$_2$ layer.}
    \label{fig:thickness}
\end{figure}

"The thickness uniformity of the wafer-stack layers was investigated using multi-point reflectometry. This allowed the tickness of the SiO$_2$ and Si layers to be determined over an area of 10's cm$^2$'s as is depicted in Fig.~\ref{fig:thickness} Here the uniformity of the wafer can be seen with the thickness of both Si layers being 202~$\pm$~1~nm (as determined from the normal distribution fitting depicted in Fig.~\ref{fig:thickness}a). Furthermore, because the silica layer thickness determines the air-gap, the uniformity of this layer is critically important. We find that this layer possesses a thickness of 57~$\pm$~3~nm. These results closely match the target thicknesses of 200~and 60~nm for the Si and SiO$_2$ layers respectively.

\subsection{Device fabrication}
\label{Sec-Device_fab}
The double-disks are fabricated using standard \ac{EBL} and \ac{RIE}. A schematic of the fabrication process is depicted in Fig.~S4. 

Electron-resist AR-P 6200.13 (CSAR, AllResist) was spincoated  onto the double-layer~SOI chips (1500 RPM), followed by 180$~^\circ\mathrm{C}$ bake producing a $\sim$580-nm coating (confirmed with thin-film reflectometry). As the total thickness of material that needs to be etched is relatively thick (2$\times$200~nm Si and 60~nm SiO$_{2}$), and the selectivity of the RIE is quite low (approaching 1:1) a thicker resist than typical SOI fabrication was found to be desirable to protect the top device layer. To compensate for the thicker resist and heterogeneous stack, electron scattering simulations and resultant \ac{PEC} were performed with GenISys Tracer and GenISys Beamer software respectively. The chips were then patterned with a Raith EBPG-5150 electron beam pattern generator (100~kV accelerating voltage, beam current 1--100~nA) for a total dosage of approximately 300~$\upmu$C/cm$^2$, followed by developing  with AR600-546 developer, O-xylene and \ac{IPA} for 90, 5, and 15 seconds respectively.

After patterning, the chips were then etched with \ac{RIE} in a three-stage process to remove the Si, SiO$_{2}$ and Si device layers sequentially. Here, the Si etches consist of SF$_6$/CHF$_3$ 20/35~sccm (82~W RF power) for 450~s, and SiO$_{2}$ etch of CHF$_3$/Ar 25/45~sccm (200~W) for 120~s.
Afterwards, any residual resist was removed by soaking the chips in Remover-PG at 60$~^\circ\mathrm{C}$ until the resist was visibly removed upon inspection with an optical microscope. The final step in the fabrication process is release of the design with \ac{HF} \ac{VPE} (at 35$~^\circ\mathrm{C}$).

\begin{figure}
    \centering
    \includegraphics[width=0.75\textwidth]{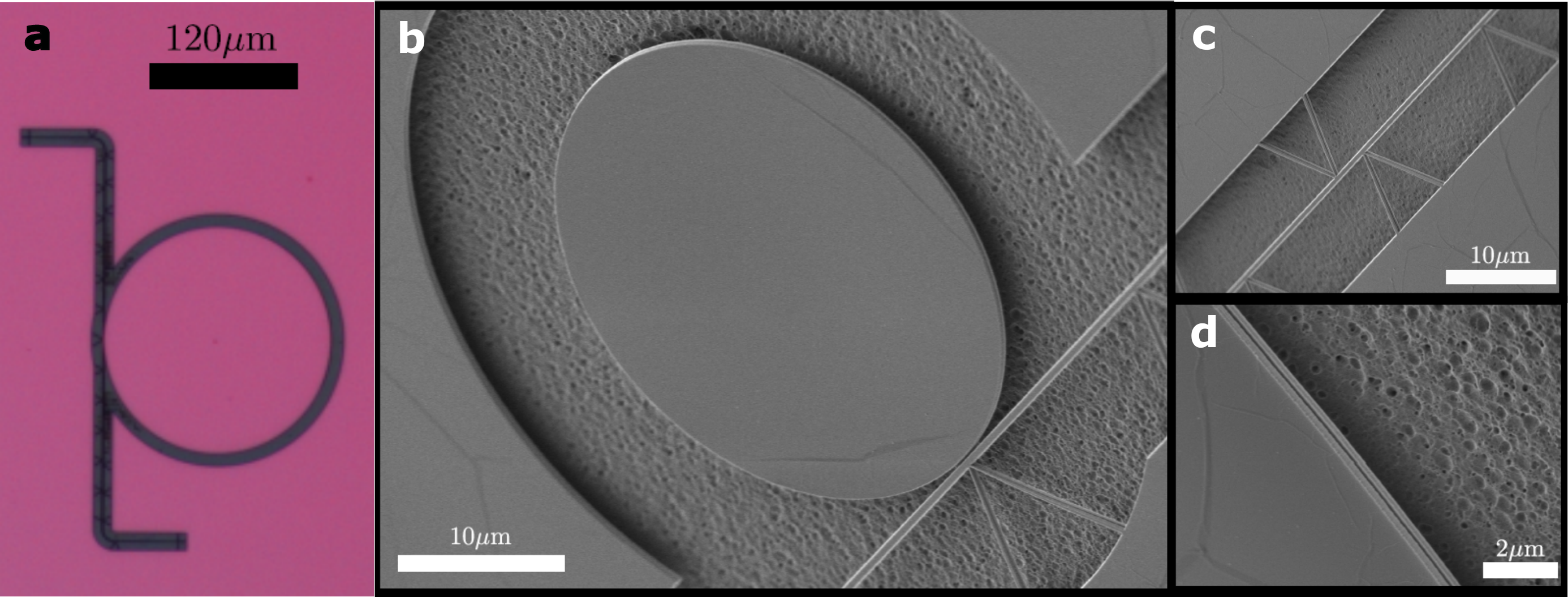}
    \caption{(a) Optical micrograph of a fabricated device (a wide-view photograph is shown in Fig.~S5).\\
    \ac{SEM} images of a (b) double-disk resonator; (c) suspended waveguide; and (d) edge of the released area.}
    \label{fig:Optical_and_SEM}
\end{figure}

The devices were inspected with optical microscopy, due to concerns surrounding degradation of their optical properties when subjected to scanning electron microscopy (SEM). Devices imaged with SEM were fabricated on a separate chip diced from the same wafer and not used for subsequent measurements. SEM pictures of a fabricated double-disk device and waveguide are included in Fig.~\ref{fig:Optical_and_SEM}(b--d). The pictures are taken at an oblique angle of $45^\circ$. The high contrast between the edges of the waveguides and disks indicate the existence of an air gap.

\section{Device Characterization}
\label{Sec:Results}
\label{Subsec:DDmeasurements}

\subsection{Optical properties of Si double-disk resonators}
\label{results:DD_opt}
\begin{figure}[thb]
     \centering
     \includegraphics[width=0.9\textwidth]{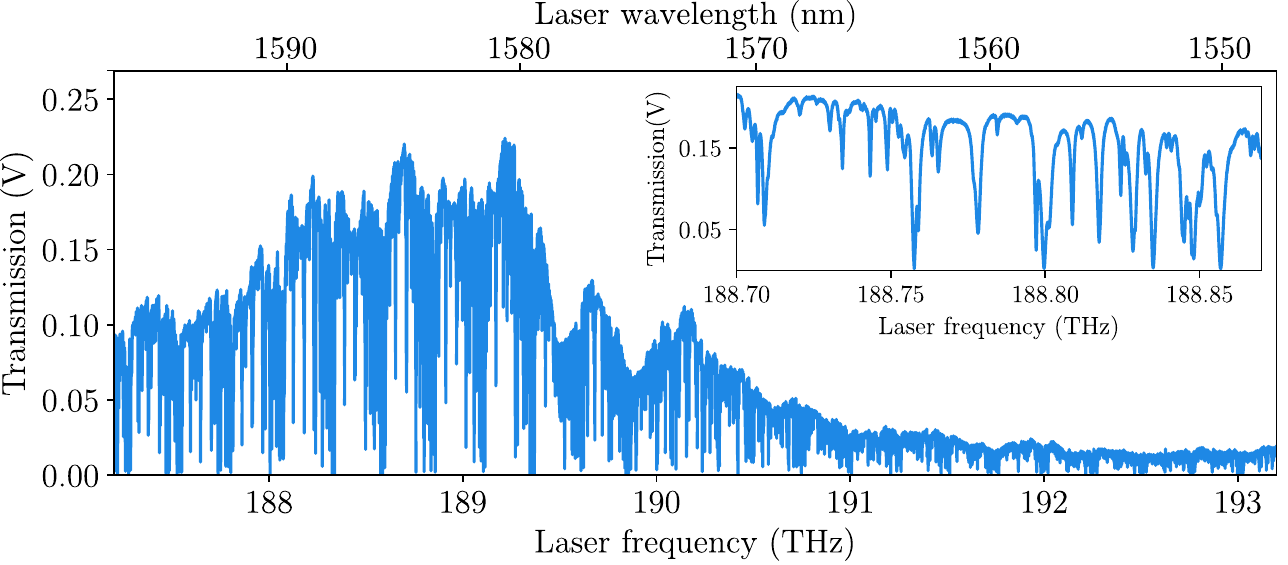}
     \caption[Transmission spectrum.]{Optical transmission spectrum of a 45$\mu$m radius disk in ambient conditions. The sharp dips are optical resonances of the double-disks. The inset shows the transmission of a narrow span (0.2~THz).}
     \label{fig:detectedPhotocurrent}
 \end{figure}

Optical transmission spectra were utilized to confirm the functionality of the waveguides and access the optical interaction with the double-disk resonators. For this, a fiber-coupled laser (EXFO T100S-HP model CU) was connected to a single sided taper, which was aligned with the waveguide using a 6-axis positioning stage. Another single sided taper was similarly aligned on the other end of the S-bend waveguide and connects to a \ac{PD} (New focus 1811-FC-AC, Newport Inc.). The \ac{PD} is connected to an oscilloscope (Tektronix DPO 3034 Digital Phosphor Oscilloscope), a computer (through a National Instruments Data Acquisition (DAQ) board) and a spectrum analyser (Liquid instruments, Moku:Pro), as depicted in Supplementary Fig.~S6. The emission wavelength of the laser was swept whilst monitoring the transmitted intensity, as is presented in Fig.~\ref{fig:detectedPhotocurrent}. Here, a high density of sharp dips are observed from the spectrum. These correspond to a \ac{FSR} of $\sim 2.8$~nm.
The \ac{FSR}  ($\Delta\lambda$) can be approximated as 
\begin{equation}
    \Delta \lambda \approx \frac{\lambda^2}{2\pi r n_{\mathrm{eff}}},
    \label{Eq.FSR}
\end{equation}
which for an $r$ = 45-$\upmu$m,  $z_g$ = 60-nm double-disk  results in an \ac{FSR} of 2.88~nm, in good agreement with experiment.

The fundamental \ac{WGM}s of the double-disks are accompanied by higher order resonances, as observed from the additional families of sharp dips with similar \ac{FSR}s. The broader features of the transmission spectrum are attributed to the polarization-dependent coupling efficiencies between the fiber tapers and on-chip waveguides \cite{Ward2014pull}. These can be compensated for by manipulating the \ac{PC}. However, it is evident from Fig.~\ref{fig:detectedPhotocurrent} (inset)  the off-resonance intensity of the transmission spectrum does not appreciably change across the frequency scales of the resonance dips. Thus these polarization effects are considered negligible over the line-width of individual resonance dips, so we will neglect them when fitting the lineshape in the following characterization.

Characterization of the optomechanical interaction was performed using a slow wavelength sweep with the devices held in vacuum (see section \ref{results:DD_mech} for details). An exemplar transmission trace is shown in Fig.~\ref{fig:TransmissionFit}. The asymmetrical lineshape is unexpected, as most \ac{WGM}s exhibit symmetrical Lorentzian profiles \cite{bowen_quantum_2015}. This asymmetry is independent of the \ac{PC}, optical power, and sweep rate, which suggests that it is not caused by thermal broadening \cite{leijssen_nonlinear_2017}. Rather, the asymmetry appears to be due to interference between the high-Q optical mode of the resonator and the continuum of states supported by the waveguide, \textit{i.e.}, a Fano resonance. The Fano-type lineshape is given by \cite{limonov2017fano}:
\begin{equation}
    \label{eq:Fano}
    T(\omega) = T_0+(T_\infty-T_0)\frac{(q-\epsilon)^2}{1+\epsilon^2},
\end{equation}
where $T_0$ and $T_\infty$ are the transmissions at resonance and far off-resonance respectively, $q$ is the Fano asymmetry parameter, and $\epsilon = 2\Delta/\kappa$ is the detuning ($\Delta = \omega_0-\omega_L$) normalised by the cavity (amplitude) loss rate, ($\kappa/2$). Here, $\omega_0$ is the cavity resonance frequency and $\omega_L$ is the laser frequency.
This lineshape was found to be in excellent agreement with the measured data (Fig.~\ref{fig:TransmissionFit}(b)). The extracted energy decay rate of the cavity is $\kappa = 2\pi\times 1.71\;\mathrm{GHz}$, which is in good agreement to the estimate based upon the the \ac{FWHM} of the transmission dip. This yields an optical Q-factor of ($\text{Q} = \omega_0/\kappa$) $\sim 2\times10^5$,
which is comparable to reports of other double-disk photonic resonators (which range from $2.5 \times 10^4$\cite{moradinejad_double-layer_2014} to $\sim10^6$\cite{jiang_high-q_2009} for Si and SiN devices respectively). The fitted Fano parameter is $q=+0.1744$.

\subsection{Optomechanical properties of Si double-disk resonators}
\label{results:DD_mech}
Initial investigations of the mechanical resonances of the disks were performed by detuning the laser around the \ac{HWHM} point (sideband detuning) of one optical mode and monitoring the \ac{PSD} of the transmitted light with a spectrum analyzer. The mechanical displacement is transduced into intensity fluctuations of the detected signal. The resulting power spectrum shows, as expected, at ambient pressure a specific kind of fluid damping known as squeeze-film damping greatly degrades the mechanical Q-factor \cite{bao2007squeeze}, resulting in mechanical Q-factors of less than 10 (see S.3.1, \& Fig.~S7). This is consistent with prior works \cite{wiederhecker2009controlling, lin2009mechanical}. Despite this, we also see the existence of a mechanical resonance at 24~MHz, as predicted by our \ac{FEM} simulation for double-disks with an undercut of $\sim$3.35$\upmu$m (Fig.~S1).

\begin{figure}[!ht]
     \centering
     \includegraphics[width=.85\textwidth]{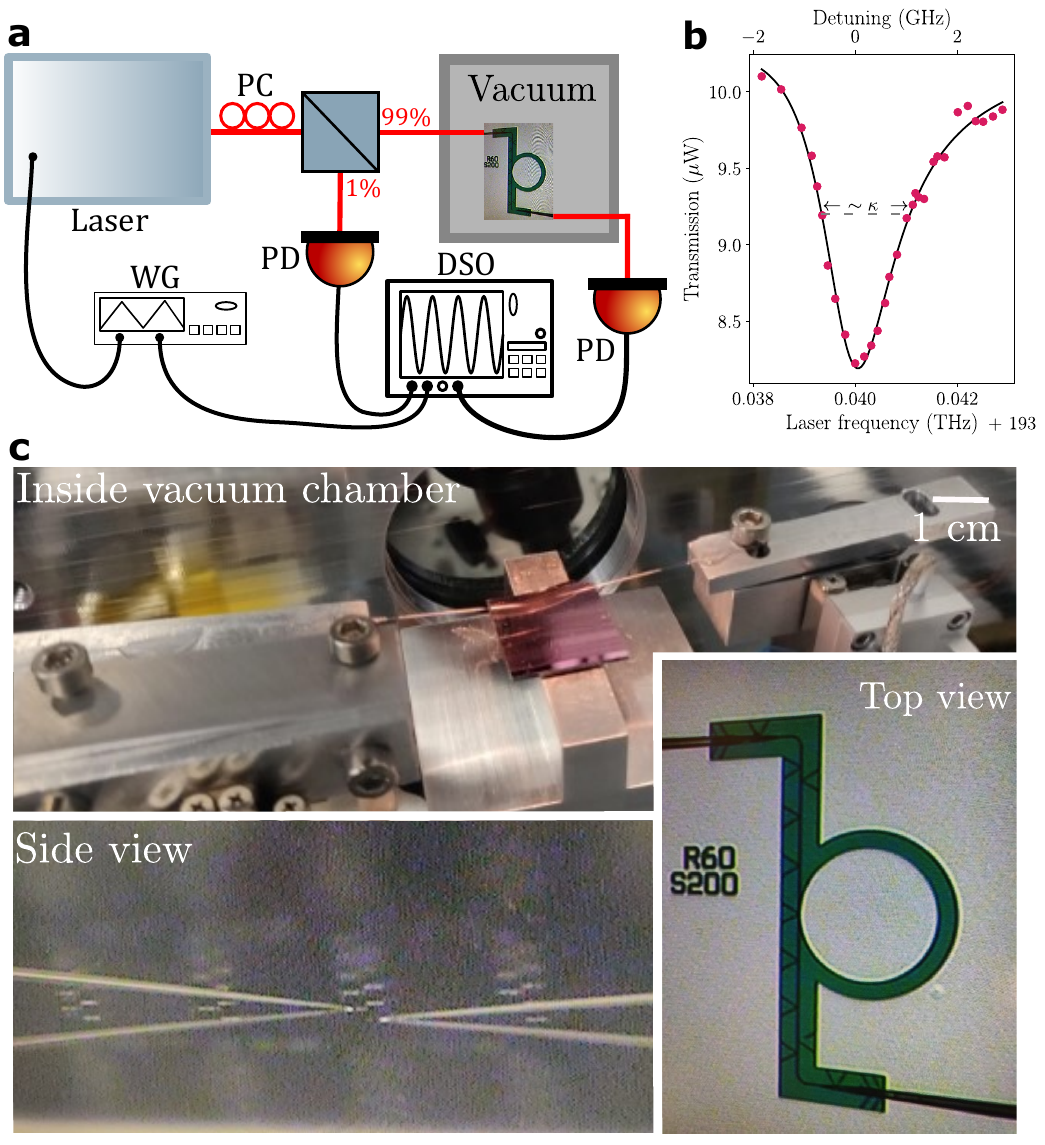}
    \caption{(a) Schematic of vacuum chamber measurement set-up for transmission measurements of silicon double-disks. A laser is coupled into a fiber. The light is guided through a polarization controller and split by a 99:1 beamsplitter; 1\% is directly connected to a power monitor, whilst the remainder is fed into the chamber and coupled to the on-chip waveguides using a single-sided tapered fiber. Transmitted light is collected by a second taper and guided to the detection photodetector (PD). The photocurrent is monitored on an oscilloscope and post-processed on a connected computer. A waveform generator is used to control the laser sweep, with the control voltage monitored at the oscilloscope to allow tracking of the output wavelength.\\
    (b) Transmission spectrum of double-disks under vacuum. The data (red dots) is fitted with a Fano lineshape, Eq.~(\ref{eq:Fano}) (black line). The resonance frequency is approximately 193.040~THz (1554.082~nm) and the Fano parameter is $q=+0.1744$.. This resonance is used for the investigations presented in section \ref{results:DD_mech}. Note that the detuning is defined so that positive detuning is to the red side of the resonance.\\
    (c) Pictures of measurement set-up for measurement of double-disks. The chip is situated on a stage, and fibers are glued onto aluminum supports mounted on linear positioners. When the vacuum chamber is closed, coupling to the disks is monitored through a top and a side window.}
     \label{fig:vacuum_chamber_setup} 
     \label{fig:setupVacuumChamberPictures}
      \label{fig:TransmissionFit}
 \end{figure}

To overcome the effects of squeeze-film damping and study the intrinsic optomechanical properties of the disks, further measurements were performed under vacuum ($\sim 10^{-6}$~mbar). The vacuum apparatus is depicted in Fig.~\ref{fig:vacuum_chamber_setup}. Here, a free-space laser (New Focus 6300 High-Power Velocity Tunable Diode Laser) is coupled into a single-mode fiber. A 99:1 beamsplitter is used to monitor the input power during the experiment prior to insertion into the chamber, and the output fiber is connected to a  \ac{PD} (New Focus 1811-FC-AC, Newport Inc.). An oscilloscope (Tektronix, DPO3034) is used for data acquisition.

Single-sided tapered fibers are inserted into the vacuum chamber using feedthroughs. The fiber tips are glued on aluminum supports and mounted on positioning stages (SmarAct Linear Stages). Two windows (top and side view) allow for monitoring of the coupling of the fiber tips to the waveguide. 
Fig.~\ref{fig:setupVacuumChamberPictures} shows the vacuum chamber experimental set-up with a mounted chip.  The microscope image presented in the Fig.~5(c) top view is the device used in the subsequent investigation of the optomechanical properties. The chamber is pumped down with a roughing-pump-backed turbomolecular pump.

\begin{figure}[!ht]
    \centering
    \includegraphics[width=\textwidth]{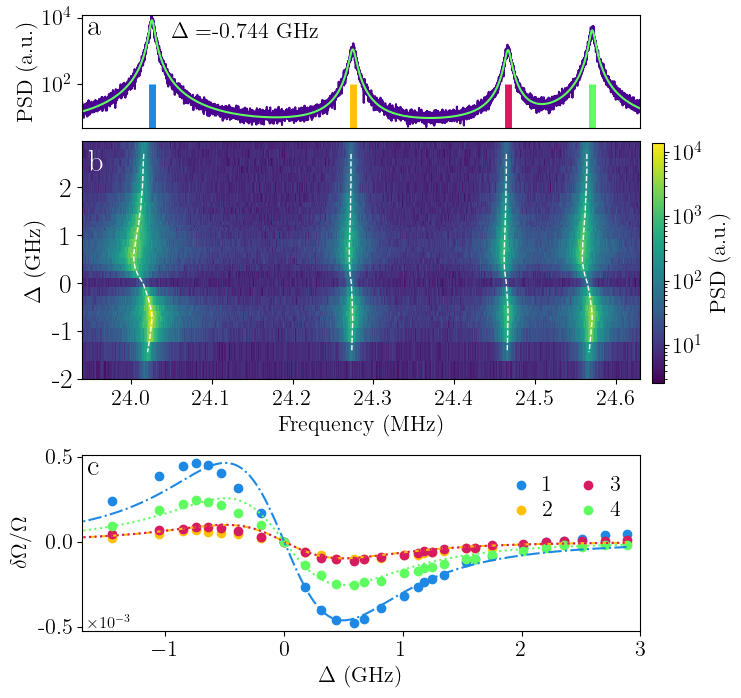}
    \caption[Power spectral densities as a function of detuning and one slice.]{(a) Sideband-detuned \ac{PSD} of the double-disk motion (blue) and fit according to Eq.~S.1 (green), with vertical bars marking the center frequencies of each resonance. Data were acquired with a detuning of $-0.744$~GHz.\\
    (b) The \ac{PSD}s as a function of optical detuning. Here the dotted white lines correspond to the centre of the fitted Lorentzian (\textit{i.e.} $\Omega_i$). \\
    (c) Optical spring effect (Eq.~\ref{Eq:SpringEffect}) frequency shifts extracted from (b). The color of each curve corresponds to those allocated in (a). Experimental points are shown as circles and fits are shown as dashed/dotted lines.}
    \label{fig:PSDSandOneFit}
\end{figure}

Fig.~\ref{fig:PSDSandOneFit}(a) depicts the \ac{PSD} of the transmitted signal when sideband-detuned (with detuning $\Delta = -0.744$~GHz). Here, what was a very broad and nondescript feature in the spectrum at ambient pressure is revealed to be four distinct mechanical modes. Fitting Lorentzian lineshapes to each peak shows that the mechanical Q-factors of the four modes are on the order of $\sim10^3$, which is comparable to previous reports on silica double-disks in vacuum conditions \cite{meng_measurement-based_2022}; this is, however, the first measurement of \textit{silicon} double-disk resonators under vacuum. The complete fit model and parameters are summarised in S.3.2 and Table~S.1. These resonances were further modelled using COMSOL to investigat higher-order azimuthal eigenmodes (\textit{i.e.} crown-shaped modes). This FEM yielded excellent agreement of the observed modes with the the m~=~0,~3,~4,~\&~5 eigenfrequencies (See Sec.~S.3.2, Table~S.2). 

Determining the optomechanical coupling rate is less straightforward, but can be achieved by exploiting the optical spring effect, wherein a detuned laser drive causes a shift of the mechanical resonance frequency \cite{bowen_quantum_2015}. The expected shift of angular frequency $\delta \Omega_i$ is given by
\begin{equation}\label{Eq:SpringEffect}
    \frac{\delta\Omega_i}{\Omega_i} =  \left[ 1 - \frac{{C_i}\left(0\right)}{Q_i} \cdot \frac{2\epsilon}{ \left(1+\epsilon^2\right)^2}\right]^{1/2}-1,
\end{equation}
where $Q_i$ is the zero-detuning mechanical Q-factor (note, we use the optomechanically shifted Q-factor in the calculation of $C$, which introduces an error on the order of 0.5\%. This is far less than the uncertainty in the fit, and therefore negligible. See S.3.2 for details.), Here $\epsilon = 2\Delta/\kappa$ and $\kappa$ is the optical linewidth (as discussed in section~\ref{results:DD_opt}). We have introduced the (resonant) optomechanical cooperativity, $C_i\left(0\right)$, which is related to the single-photon optomechanical coupling rate by
\begin{equation}
    g_0 = \sqrt{\frac{C_i\left(0\right) \kappa \Gamma_i}{4n\left(0\right)}},
\end{equation}
where $n\left(\Delta\right)$ is the number of intracavity photons when the cavity is pumped at detuning $\Delta$. Thus, we can recover $g_0$ by fitting multiple mechanical spectra recorded at different detunings.

Detuning-dependent power spectra are shown in Fig.~\ref{fig:PSDSandOneFit}(b). The corresponding optical spring effect fittings (Eq.~\ref{Eq:SpringEffect}) are shown in Fig.~\ref{fig:PSDSandOneFit}(c), with numerical values quoted in Table~\ref{table:SpringFitParams}. Two of these modes (1 and 4) achieve cooperativities greater than unity, which is an important threshold for onset of instability in optomechanical amplification \cite{bowen_quantum_2015}.

To convert $C_i\left(0\right)$ to $g_0$ we require the resonant intracavity photon number ($n\left(0\right) \simeq 5 \times 10^4$), which can be calculated from the approximate waveguide photon number and the coupling strength into the cavity (see S.3.3 \& S.3.4 for details). We find that the optomechanical coupling rates lie between $g_0/2\pi = 7.78$~kHz and $g_0/2\pi = 15.47$~kHz for the modes presented in this study (see Table~\ref{table:SpringFitParams}). The discrepancies between the modes are likely due to fabrication inconsistencies and minor ellipticity in the disk geometry. These values are consistent with reports for hybridized silica \ac{WGM} resonators such as those presented in \cite{rosenberg2009static, meng_measurement-based_2022}. Note that we were unable to observe optical resonances in our low-radius devices (likely due to fabrication imperfections), which were expected to have the largest optomechanical coupling rates. Our devices can therefore be further optimized to improve these $g_0$ values beyond those obtained in silica. The maximum experimentally determined value of $g_0$ is $\approx 1/8$ of the expected valued based on the FEM (Sec.~(\ref{results:FEM})), which is reasonably congruent. The disparity likely resulting from fabrication inconsistencies or imperfections.
 
\begin{table}[tbh]
\centering
\begin{tabular}{|c|c|c|c|c|}\hline
Mode number & $C(0)$ & $C_0\times10^{+5}$ & $g_0/2\pi$~(kHz) 
    \\ \hline \hline
    1 & $4.11~(\pm~0.15)$ & $9.05~(\pm~0.32)$ & $15.47~(\pm~0.12)$\\
    2 & $0.82~(\pm~0.06)$ & $1.80~(\pm~0.13)$ & $7.86~(\pm~0.08)$\\
    3 & $0.87~(\pm~0.17)$ & $1.92~(\pm~0.37)$ & $7.78~(\pm~0.13)$\\
    4 & $2.31~(\pm~0.07)$ & $5.08~(\pm~0.15)$ & $12.18~(\pm~0.08)$\\
    \hline
\end{tabular}
\caption{Cooperativities and coupling rates of the investigated mechanical modes extracted from the curve fitting presented in Fig.~\ref{fig:PSDSandOneFit}. All of the uncertainties presented here derive from the fitting of experimental data to Eq.~(\ref{Eq:SpringEffect}).}
\label{table:SpringFitParams}
\end{table}

\section{Conclusion}
We have demonstrated the first wafer-scale fabrication of double-layer silicon-on-insulator wafers. These custom wafers were then utilized in the fabrication of freestanding Si double-disk \ac{WGM} optomechanical resonators with associated on-chip silicon waveguides, enabling integration with an enormous variety of existing Si-based technologies. 

Further, we presented the first observation of the optomechanical spring-effect in silicon double-disks. Four mechanical modes were identified that show a strong change in mechanical resonance frequency as a function of detuning. The optomechanical cooperativity for each mode is extracted from fitting to the experimental data. Using this, we determine the single-photon coupling rates in these devices.

Future devices could be improved by increasing the undercut between the disks and reducing the disks' diameters. Increasing the undercut makes the mechanical modes less sensitive to fabrication imperfections, and also tends to improve the mechanical quality factor. Reducing the disk radius is also predicted to lead to higher optomechanical coupling rates.

Further research with these devices could involve the integration of these devices with a wide range of applications including single-photon defect emitters, opto-electronics or microwave resonators which could prove useful for the fields of quantum computing and quantum state transfer. 





\begin{backmatter}
\bmsection{Funding}
Defence Science and Technology Group (ASCA); Australian Research Council (CE170100009, CE230100021): Air Force Office of Scientific Research (FA9550-20-1-0391, FA9550-22-1-0047).

\bmsection{Acknowledgments}
The authors acknowledge the facilities, and the scientific and technical assistance, of the Australian Microscopy \& Microanalysis Research Facility at the Centre for Microscopy and Microanalysis, The University of Queensland. This work was performed in part at the Queensland node of the Australian National Fabrication Facility, a company established under the National Collaborative Research Infrastructure Strategy to provide nano- and microfabrication facilities for Australia's researchers. This work was supported by the Australian Government through the Next Generation Technologies Fund (now managed through ASCA). This work was financially supported by the Australian Research Council (ARC) Centres of excellence for Engineered Quantum systems (EQUS, Grant No. CE170100009) and Quantum Biotechnology (Grant No. CE230100021). This work was supported by the Air Force Office of Scientific Research (AFOSR, FA9550-20-1-0391 \& FA9550-22-1-0047).

\bmsection{Disclosures}
The authors declare no conflicts of interest.

\bmsection{Data availability}
Data underlying the results presented in this paper are not publicly available at this time but may be obtained from the authors upon reasonable request.

\bmsection{Supplemental document}
See Supplemental Material for supporting content, including further FEM modeling, greater detail about the double-layer~SOI wafer production process and the device fabrication, results on optomechanics at ambient conditions, and investigations into the cavity photon number and photon coupling rates.

\end{backmatter}


\bibliography{Main}

\begin{thebibliography}{10}
\newcommand{\enquote}[1]{``#1''}

\bibitem{Frustaci_WGM_2019}
S.~Frustaci and F.~Vollmer, \enquote{Whispering-gallery mode (wgm) sensors:
  review of established and wgm-based techniques to study protein
  conformational dynamics,} {\protect\JournalTitle{Current Opinion in Chemical
  Biology}} \textbf{51}, 66--73 (2019).

\bibitem{Liu_sensors_2021}
X.~Liu, W.~Liu, Z.~Ren, Y.~Ma, B.~Dong, G.~Zhou, and C.~Lee, \enquote{Progress
  of optomechanical micro/nano sensors: a review,}
  {\protect\JournalTitle{International Journal of Optomechatronics}}
  \textbf{15}, 120--159 (2021).

\bibitem{Gotardo_23}
F.~Gotardo, B.~J. Carey, H.~Greenall, G.~I. Harris, E.~Romero, D.~Bulla, E.~M.
  Bridge, J.~S. Bennett, S.~Foster, and W.~P. Bowen,
  \enquote{Waveguide-integrated chip-scale optomechanical magnetometer,}
  {\protect\JournalTitle{Opt. Express}} \textbf{31}, 37663--37672 (2023).

\bibitem{basiri-esfahani_precision_2019}
S.~Basiri-Esfahani, A.~Armin, S.~Forstner, and W.~P. Bowen, \enquote{Precision
  ultrasound sensing on a chip,} {\protect\JournalTitle{Nature Communications}}
  \textbf{10}, 132 (2019).

\bibitem{kristensen2023longlived}
M.~B. Kristensen, N.~Kralj, E.~Langman, and A.~Schliesser, \enquote{A
  long-lived and efficient optomechanical memory for light,}  (2023).

\bibitem{lei_fully_2022}
Y.~Lei, Z.-G. Hu, M.~Wang, Y.-M. Gao, Z.~Zuo, X.~Xu, and B.-B. Li,
  \enquote{Fully reconfigurable optomechanical add-drop filters,}
  {\protect\JournalTitle{Applied Physics Letters}} \textbf{121}, 181110 (2022).

\bibitem{chan_optical_2009}
J.~Chan, M.~Eichenfield, R.~Camacho, and O.~Painter, \enquote{Optical and
  mechanical design of a “zipper” photonic crystal optomechanical cavity,}
  {\protect\JournalTitle{Optics Express}} \textbf{17}, 3802 (2009).

\bibitem{bowen_quantum_2015}
W.~P. Bowen and G.~J. Milburn, \emph{Quantum optomechanics} (CRC Press, 2015).

\bibitem{Tang_entanglment_2022}
J.-D. Tang, Q.-Z. Cai, Z.-D. Cheng, N.~Xu, G.-Y. Peng, P.-Q. Chen, D.-G. Wang,
  Z.-W. Xia, Y.~Wang, H.-Z. Song, Q.~Zhou, and G.-W. Deng, \enquote{A
  perspective on quantum entanglement in optomechanical systems,}
  {\protect\JournalTitle{Physics Letters A}} \textbf{429}, 127966 (2022).

\bibitem{mancini_optomechanical_1998}
S.~Mancini, D.~Vitali, and P.~Tombesi, \enquote{Optomechanical {Cooling} of a
  {Macroscopic} {Oscillator} by {Homodyne} {Feedback},}
  {\protect\JournalTitle{Physical Review Letters}} \textbf{80}, 688--691
  (1998).

\bibitem{aspelmeyer_cavity_2014}
M.~Aspelmeyer, T.~J. Kippenberg, and F.~Marquardt, \enquote{Cavity
  optomechanics,} {\protect\JournalTitle{Reviews of Modern Physics}}
  \textbf{86}, 1391--1452 (2014).

\bibitem{ma_radiation-pressure-driven_2007}
R.~Ma, A.~Schliesser, P.~Del'Haye, A.~Dabirian, G.~Anetsberger, and T.~J.
  Kippenberg, \enquote{Radiation-pressure-driven vibrational modes in
  ultrahigh-{Q} silica microspheres,} {\protect\JournalTitle{Optics Letters}}
  \textbf{32}, 2200 (2007).

\bibitem{shen_reconfigurable_2018}
Z.~Shen, Y.-L. Zhang, Y.~Chen, F.-W. Sun, X.-B. Zou, G.-C. Guo, C.-L. Zou, and
  C.-H. Dong, \enquote{Reconfigurable optomechanical circulator and directional
  amplifier,} {\protect\JournalTitle{Nature Communications}} \textbf{9}, 1797
  (2018).

\bibitem{lee_cooling_2010}
K.~H. Lee, T.~G. McRae, G.~I. Harris, J.~Knittel, and W.~P. Bowen,
  \enquote{Cooling and {Control} of a {Cavity} {Optoelectromechanical}
  {System},} {\protect\JournalTitle{Physical Review Letters}} \textbf{104},
  123604 (2010).

\bibitem{verhagen_quantum-coherent_2012}
E.~Verhagen, S.~Deléglise, S.~Weis, A.~Schliesser, and T.~J. Kippenberg,
  \enquote{Quantum-coherent coupling of a mechanical oscillator to an optical
  cavity mode,} {\protect\JournalTitle{Nature}} \textbf{482}, 63--67 (2012).

\bibitem{ruesink_optical_2018}
F.~Ruesink, J.~P. Mathew, M.-A. Miri, A.~Alù, and E.~Verhagen,
  \enquote{Optical circulation in a multimode optomechanical resonator,}
  {\protect\JournalTitle{Nature Communications}} \textbf{9}, 1798 (2018).

\bibitem{bekker_free_2018}
C.~Bekker, C.~G. Baker, R.~Kalra, H.-H. Cheng, B.-B. Li, V.~Prakash, and W.~P.
  Bowen, \enquote{Free spectral range electrical tuning of a high quality
  on-chip microcavity,} {\protect\JournalTitle{Optics Express}} \textbf{26},
  33649 (2018).

\bibitem{bekker2017injection}
C.~Bekker, R.~Kalra, C.~Baker, and W.~P. Bowen, \enquote{Injection locking of
  an electro-optomechanical device,} {\protect\JournalTitle{Optica}}
  \textbf{4}, 1196--1204 (2017).

\bibitem{lu_high-frequency_2015}
X.~Lu, J.~Y. Lee, and Q.~Lin, \enquote{High-frequency and high-quality silicon
  carbide optomechanical microresonators,} {\protect\JournalTitle{Scientific
  Reports}} \textbf{5}, 17005 (2015).

\bibitem{jiang_chip-scale_2016}
W.~C. Jiang and Q.~Lin, \enquote{Chip-scale cavity optomechanics in lithium
  niobate,} {\protect\JournalTitle{Scientific Reports}} \textbf{6}, 36920
  (2016).

\bibitem{jiang_high-q_2009}
X.~Jiang, Q.~Lin, J.~Rosenberg, K.~Vahala, and O.~Painter, \enquote{High-{Q}
  double-disk microcavities for cavity optomechanics,}
  {\protect\JournalTitle{Optics Express}} \textbf{17}, 20911 (2009).

\bibitem{meng_measurement-based_2022}
C.~Meng, G.~A. Brawley, S.~Khademi, E.~M. Bridge, J.~S. Bennett, and W.~P.
  Bowen, \enquote{Measurement-based preparation of multimode mechanical
  states,} {\protect\JournalTitle{Science Advances}} \textbf{8}, 7585 (2022).

\bibitem{ward2011wgm}
J.~Ward and O.~Benson, \enquote{Wgm microresonators: sensing, lasing and
  fundamental optics with microspheres,}  (2011).

\bibitem{chan_laser_2011}
J.~Chan, T.~P.~M. Alegre, A.~H. Safavi-Naeini, J.~T. Hill, A.~Krause,
  S.~Gröblacher, M.~Aspelmeyer, and O.~Painter, \enquote{Laser cooling of a
  nanomechanical oscillator into its quantum ground state,}
  {\protect\JournalTitle{Nature}} \textbf{478}, 89--92 (2011).

\bibitem{gavartin_optomechanical_2011}
E.~Gavartin, R.~Braive, I.~Sagnes, O.~Arcizet, A.~Beveratos, T.~J. Kippenberg,
  and I.~Robert-Philip, \enquote{Optomechanical {Coupling} in a
  {Two}-{Dimensional} {Photonic} {Crystal} {Defect} {Cavity},}
  {\protect\JournalTitle{Physical Review Letters}} \textbf{106}, 203902 (2011).

\bibitem{leijssen_nonlinear_2017}
R.~Leijssen, G.~R. La~Gala, L.~Freisem, J.~T. Muhonen, and E.~Verhagen,
  \enquote{Nonlinear cavity optomechanics with nanomechanical thermal
  fluctuations,} {\protect\JournalTitle{Nature Communications}} \textbf{8},
  ncomms16024 (2017).

\bibitem{ren_two-dimensional_2020}
H.~Ren, M.~H. Matheny, G.~S. MacCabe, J.~Luo, H.~Pfeifer, M.~Mirhosseini, and
  O.~Painter, \enquote{Two-dimensional optomechanical crystal cavity with high
  quantum cooperativity,} {\protect\JournalTitle{Nature Communications}}
  \textbf{11}, 3373 (2020).

\bibitem{lu_silicon_2020}
X.~Lu, J.~Y. Lee, and Q.~Lin, \enquote{Silicon carbide zipper photonic crystal
  optomechanical cavities,} {\protect\JournalTitle{Applied Physics Letters}}
  \textbf{116}, 221104 (2020).

\bibitem{lin2009mechanical}
Q.~Lin, J.~Rosenberg, X.~Jiang, K.~J. Vahala, and O.~Painter,
  \enquote{Mechanical oscillation and cooling actuated by the optical gradient
  force,} {\protect\JournalTitle{Physical review letters}} \textbf{103}, 103601
  (2009).

\bibitem{lee2010silicon}
S.~Lee, S.~C. Eom, J.~S. Chang, C.~Huh, G.~Y. Sung, and J.~H. Shin, \enquote{A
  silicon nitride microdisk resonator with a 40-nm-thin horizontal air slot,}
  {\protect\JournalTitle{Optics express}} \textbf{18}, 11209--11215 (2010).

\bibitem{wiederhecker2011broadband}
G.~S. Wiederhecker, S.~Manipatruni, S.~Lee, and M.~Lipson, \enquote{Broadband
  tuning of optomechanical cavities,} {\protect\JournalTitle{Optics Express}}
  \textbf{19}, 2782--2790 (2011).

\bibitem{zheng2019high}
Y.~Zheng, Z.~Fang, S.~Liu, Y.~Cheng, and X.~Chen, \enquote{High-q exterior
  whispering-gallery modes in a double-layer crystalline microdisk resonator,}
  {\protect\JournalTitle{Physical Review Letters}} \textbf{122}, 253902 (2019).

\bibitem{moradinejad_double-layer_2014}
H.~Moradinejad, A.~H. Atabaki, A.~H. Hosseinnia, A.~A. Eftekhar, and A.~Adibi,
  \enquote{Double-{Layer} {Crystalline} {Silicon} on {Insulator} {Material}
  {Platform} for {Integrated} {Photonic} {Applications},}
  {\protect\JournalTitle{IEEE Photonics Journal}} \textbf{6}, 1--8 (2014).

\bibitem{espinel_brillouin_2017}
Y.~A.~V. Espinel, F.~G.~S. Santos, G.~O. Luiz, T.~P.~M. Alegre, and G.~S.
  Wiederhecker, \enquote{Brillouin {Optomechanics} in {Coupled} {Silicon}
  {Microcavities},} {\protect\JournalTitle{Scientific Reports}} \textbf{7},
  43423 (2017).

\bibitem{dehghannasiri_integrated_2018}
R.~Dehghannasiri, H.~Moradinejad, T.~Fan, A.~H. Hosseinnia, A.~A. Eftekhar, and
  A.~Adibi, \enquote{Integrated {Optomechanical} {Resonators} in
  {Double}-{Layer} {Crystalline} {Silicon} {Platforms},} in \emph{2018 {IEEE}
  {Photonics} {Conference} ({IPC}),}  (IEEE, Reston, VA, 2018), pp. 1--2.

\bibitem{Hollenbach:2020-defect}
M.~Hollenbach, Y.~Berenc\'{e}n, U.~Kentsch, M.~Helm, and G.~V. Astakhov,
  \enquote{Engineering telecom single-photon emitters in silicon for scalable
  quantum photonics,} {\protect\JournalTitle{Opt. Express}} \textbf{28},
  26111--26121 (2020).

\bibitem{Durand-2021_singe-photon-defect}
A.~Durand, Y.~Baron, W.~Redjem, T.~Herzig, A.~Benali, S.~Pezzagna, J.~Meijer,
  A.~Y. Kuznetsov, J.-M. G\'erard, I.~Robert-Philip, M.~Abbarchi, V.~Jacques,
  G.~Cassabois, and A.~Dr\'eau, \enquote{Broad diversity of near-infrared
  single-photon emitters in silicon,} {\protect\JournalTitle{Phys. Rev. Lett.}}
  \textbf{126}, 083602 (2021).

\bibitem{briggs_wafer-bonded_2009}
R.~M. Briggs, M.~Shearn, A.~Scherer, and H.~A. Atwater, \enquote{Wafer-bonded
  single-crystal silicon slot waveguides and ring resonators,}
  {\protect\JournalTitle{Applied Physics Letters}} \textbf{94}, 021106 (2009).

\bibitem{ohke1995new}
S.~Ohke, Y.~Satomura, T.~Umeda, and Y.~Cho, \enquote{A new integral expression
  for the effective refractive index of optical waveguides and its application
  for a fast numerical solution finder for the effective refractive index,}
  {\protect\JournalTitle{Optics communications}} \textbf{118}, 227--234 (1995).

\bibitem{Xia_Optomech_review_2020}
J.~Xia, Q.~Qiao, G.~Zhou, F.~S. Chau, and G.~Zhou, \enquote{Opto-mechanical
  photonic crystal cavities for sensing application,}
  {\protect\JournalTitle{Applied Sciences}} \textbf{10} (2020).

\bibitem{lauk_perspectives_2020}
N.~Lauk, N.~Sinclair, S.~Barzanjeh, J.~P. Covey, M.~Saffman, M.~Spiropulu, and
  C.~Simon, \enquote{Perspectives on quantum transduction,}
  {\protect\JournalTitle{Quantum Science and Technology}} \textbf{5}, 020501
  (2020).

\bibitem{rodriguez2011designing}
A.~W. Rodriguez, D.~Woolf, P.-C. Hui, E.~Iwase, A.~P. McCauley, F.~Capasso,
  M.~Loncar, and S.~G. Johnson, \enquote{Designing evanescent optical
  interactions to control the expression of casimir forces in optomechanical
  structures,} {\protect\JournalTitle{Applied Physics Letters}} \textbf{98}
  (2011).

\bibitem{rodriguez2011bonding}
A.~W. Rodriguez, A.~P. McCauley, P.-C. Hui, D.~Woolf, E.~Iwase, F.~Capasso,
  M.~Loncar, and S.~G. Johnson, \enquote{Bonding, antibonding and tunable
  optical forces in asymmetric membranes,} {\protect\JournalTitle{Optics
  express}} \textbf{19}, 2225--2241 (2011).

\bibitem{Winger_2011}
M.~Winger, T.~D. Blasius, T.~P.~M. Alegre, A.~H. Safavi-Naeini, S.~Meenehan,
  J.~Cohen, S.~Stobbe, and O.~Painter, \enquote{A chip-scale integrated
  cavity-electro-optomechanics platform,} {\protect\JournalTitle{Opt. Express}}
  \textbf{19}, 24905--24921 (2011).

\bibitem{woolf2013optomechanical}
D.~Woolf, P.-C. Hui, E.~Iwase, M.~Khan, A.~W. Rodriguez, P.~Deotare, I.~Bulu,
  S.~G. Johnson, F.~Capasso, and M.~Loncar, \enquote{Optomechanical and
  photothermal interactions in suspended photonic crystal membranes,}
  {\protect\JournalTitle{Optics express}} \textbf{21}, 7258--7275 (2013).

\bibitem{buck_joseph_robert_cavity_2003}
J.~R. Buck, \enquote{Cavity {QED} in microsphere and {Fabry}-{Perot} cavities,}
  Ph.D. thesis, California Institute of Technology (2003). Medium: PDF Version
  Number: Final.

\bibitem{Tiecke2015}
T.~G. Tiecke, K.~P. Nayak, J.~D. Thompson, T.~Peyronel, N.~P. de~Leon,
  V.~Vuleti\'{c}, and M.~D. Lukin, \enquote{Efficient fiber-optical interface
  for nanophotonic devices,} {\protect\JournalTitle{Optica}} \textbf{2}, 70--75
  (2015).

\bibitem{moradinejad_hybrid_2017}
H.~Moradinejad, \enquote{Hybrid multi-layer {CMOS}-compatible material platform
  for high-performance integrated nanophotonics,} Ph.D. thesis, Georgia
  Institute of Technology (2017).

\bibitem{zhang_fundamentals_2019}
J.~X. Zhang and K.~Hoshino, \enquote{Fundamentals of nano/microfabrication and
  scale effect,} in \emph{Molecular {Sensors} and {Nanodevices},}  (Elsevier,
  2019), pp. 43--111.

\bibitem{moriceau_overview_2011}
H.~Moriceau, F.~Rieutord, F.~Fournel, Y.~Le~Tiec, L.~Di~Cioccio, C.~Morales,
  A.~M. Charvet, and C.~Deguet, \enquote{Overview of recent direct wafer
  bonding advances and applications,} {\protect\JournalTitle{Advances in
  Natural Sciences: Nanoscience and Nanotechnology}} \textbf{1}, 043004 (2011).

\bibitem{pantzas_measuring_2020}
K.~Pantzas, F.~Fournel, A.~Talneau, G.~Patriarche, and E.~Le~Bourhis,
  \enquote{Measuring the surface bonding energy: {A} comparison between the
  classical double-cantilever beam experiment and its nanoscale analog,}
  {\protect\JournalTitle{AIP Advances}} \textbf{10}, 045006 (2020).

\bibitem{Ward2014pull}
J.~M. Ward, A.~Maimaiti, V.~H. Le, and S.~N. Chormaic, \enquote{{Contributed
  Review: Optical micro- and nanofiber pulling rig},}
  {\protect\JournalTitle{Review of Scientific Instruments}} \textbf{85} (2014).
  111501.

\bibitem{limonov2017fano}
M.~F. Limonov, M.~V. Rybin, A.~N. Poddubny, and Y.~S. Kivshar, \enquote{Fano
  resonances in photonics,} {\protect\JournalTitle{Nature Photonics}}
  \textbf{11}, 543--554 (2017).

\bibitem{bao2007squeeze}
M.~Bao and H.~Yang, \enquote{Squeeze film air damping in mems,}
  {\protect\JournalTitle{Sensors and Actuators A: Physical}} \textbf{136},
  3--27 (2007).

\bibitem{wiederhecker2009controlling}
G.~S. Wiederhecker, L.~Chen, A.~Gondarenko, and M.~Lipson, \enquote{Controlling
  photonic structures using optical forces,} {\protect\JournalTitle{Nature}}
  \textbf{462}, 633--636 (2009).

\bibitem{rosenberg2009static}
J.~Rosenberg, Q.~Lin, and O.~Painter, \enquote{Static and dynamic wavelength
  routing via the gradient optical force,} {\protect\JournalTitle{Nature
  Photonics}} \textbf{3}, 478--483 (2009).

\end{thebibliography}

\end{document}